\documentclass{ws-procs961x669}            
\begin{document}
\title{Varying fundamental constants and dark energy in the ESPRESSO era}

\author{C. J. A. P. Martins$^*$}

\address{Centro de Astrof\'{\i}sica da Universidade do Porto, and\\
Instituto de Astrof\'{\i}sica e Ci\^encias do Espa\c co, Universidade do Porto,\\
Rua das Estrelas, 4150-762 Porto, Portugal\\
$^*$E-mail: Carlos.Martins@astro.up.pt}

\begin{abstract}
The observational evidence for the recent acceleration of the universe shows that canonical theories of cosmology and particle physics are incomplete and that new physics is out there, waiting to be discovered. A compelling task for astrophysical facilities is to search for, identify and ultimately characterize this new physics. I present very recent developments in tests of the stability of nature's fundamental constants, as well as their impact on physics paradigms beyond the standard model. Specifically I discuss new observational constraints at low redshifts and at the BBN epoch, and highlight their different implications for canonical quintessence-type models and for non-canonical string-theory inspired models. Finally I also present new forecasts, based on realistic simulated data, of the gains in sensitivity for these constraints expected from ELT-HIRES, on its own and in combination with Euclid.
\end{abstract}


\bodymatter

\section{Introduction}
The observational evidence for the acceleration of the universe shows that our canonical theories of cosmology and particle physics are at least incomplete, and possibly incorrect. Is dark energy a cosmological constant (i.e. vacuum energy)? If the answer is yes, it is ten to some large power times smaller than our Quantum Field Theory based expectations. If the answer is no, then the Einstein Equivalence Principle must be violated. Either way, new physics is out there, waiting to be discovered; we must search for, identify and characterize this new physics.

The CosmoESPRESSO team uses the universe as a laboratory to address, with precision spectroscopy and other observational, computational and theoretical tools, 6 grand-challenge questions:

\begin{arabiclist}[(6)]
\item[(1)] Are the laws of physics universal?
\item[(2)] Is gravity just geometry (i.e., a fictitious force)?
\item[(3)] What makes the universe accelerate?
\item[(4)] Can we find fossil relics of the early universe? 
\item[(5)] How do we optimize next-generation facilities?
\item[(6)] How do we prepare next-generation (astro)physicists?
\end{arabiclist}

In what follows I will highlight recent contributions of the CosmoESPRESSO team to this fundamental quest, pertaining to fundamental constants.

\section{Scalars, because they are there}

We know since 2012 (thanks to the LHC) that fundamental scalar fields are among Nature's building blocks. Even before this discovery they were already widely used in cosmology, e.g., to describe inflation, cosmic defects, dynamical dark energy, and dynamical fundamental couplings. We also expect that cosmological scalar fields will naturally couple to the rest of the model, leading to long-range forces and ‘varying constants’ \cite{Dicke,Carroll,Damour}. Of particular interest in what follows are electromagnetic sector couplings, which yield spacetime variations of the fine-structure constant, with multiple testable fingerprints. A recent theoretical and observational overview of the field can be found in Ref. \citenum{ROPP}.

One of the most actively pursued test of the stability of fundamental couplings consists of high resolution astrophysical spectroscopy measurements of the fine-structure constant $\alpha$, the proton to electron mass ratio $\mu$, the proton gyromagnetic ratio $g_p$, or combinations thereof. A recent joint likelihood analysis of all currently available data \cite{Meritxell} leads to a very mild (one to two standard deviations of statistical significance) preference for astrophysical variations (as compared to the local laboratory values), at the parts per million level of relative variation, All this data spans the redshift range $0.2<z<4.2$. However, it is also known that, at least in the case of $\alpha$ measurements which make up most of the full dataset, there are systematics at the level of at least parts per million. A new generation of more robust measurements is therefore necessary.

\section{Aiming higher}

In the most natural and physically realistic models for time (redshift) variations of fundamental couplings, such variations are monotonic. Moreover, one expects that the putative scalar field driving this variation will be significantly damped at the onset of the dark energy domination phase of the universe (i.e., the recent acceleration phase). Both of these are theoretical motivations for testing the stability of fundamental constants such as $\alpha$, though direct measurements, at higher redshifts---in other words, deep into the matter era, where any relative variations are expected to be larger.

An example of work towards this goal relies on using observations of the redshift $z=7.085$ quasar J1120+0641 \cite{Mortlock} to constrain variations of  $\alpha$ over the redshift range $z=5.51$ to $z=7.06$. These led to the four highest redshift direct measurements of $\alpha$ (the previous highest redshift direct measurement was at $z=4.18$), with the latter corresponding to a look-back time 12.96 Gyr (for the current best-fit cosmological parameters in the standard $\Lambda$CDM model) . A total of about 30 hours of data from the X-SHOOTER spectrograph on the European Southern Observatory's Very Large Telescope (VLT) was used, which also makes this the first direct measurement of $\alpha$ in the infrared. Finally, the analysis relied on a new AI-based method, aiming to remove some possible dependency of the final result on human-made choices. The weighted mean strength of the electromagnetic force over this redshift range in this location in the universe, reported in Ref. \citenum{Wilczynska},is measured to be
\begin{equation}
\frac{\Delta\alpha}{\alpha} = \frac{\alpha(z) - \alpha_0}{\alpha_0} = (-2.18\pm 7.27) \times 10^{-5}\,, 
\end{equation}
where $\alpha_0$ denotes the local laboratory value, i.e. we find no evidence for a time variation. The sensitivity of the combined measurement is only at the tens of parts per million level, as compared to the parts per million (nominal) sensitivity of optical measurements at lower redshifts, but this first result should primarily be seen as a proof of concept analysis, demonstrating that such high redshift measurements can indeed be done.

As for measurements in the optical, the arrival of the ESPRESSO spectrograph \cite{ESPRESSO1}, operating at the combined Coud\'e focus of the VLT, enables new and more stringent tests. Preliminary analyses already demonstrate that the dominant sources of systematics limiting previous high resolution spectrographs are not present \cite{ESPRESSO2}, and the first results of its measurements of the fine-structure constant should be reported soon.

Broadly speaking, the direct impact of ESPRESSO on cosmology and fundamental physics will come from at least five different types of observations:
\begin{arabiclist}[(5)]
\item[(1)] Direct measurements of the fine-structure constant $\alpha$, at redshifts between $z\sim1$ and $z\sim4$, using various ions
\item[(2)] Direct measurements of the proton to electron mass ratio $\mu$, at redshifts between $z\sim2$ and $z\sim4$, using molecular Hydrogen and Carbon Monoxide
\item[(3)] Direct measurements of the cosmic microwave background temperature at redshifts, at redshifts between $z\sim2$ and $z\sim4$, using Carbon Monoxide and neutral Carbon
\item[(4)] New measurements of the primordial Deuterium abundance (of relevance for BBN, as discussed in what follows)
\item[(5)] Additional probes, including deep spectra, lensed QSOs and precursor redshift drift measurements.
\end{arabiclist}
In addition to their direct fundamental relevance, these will also impact the quest to characterize dark energy properties \cite{Leite}.

\section{BBN with GUTs}

Big Bang Nucleosynthesis is one of the cornerstones of the Hot Big Bang model, but its success has been limited by the long-standing Lithium problem \cite{Fields}; more recently, a possible Deuterium discrepancy has also been suggested \cite{Pitrou20}.

If a fiducial theoretical model is chosen and the relevant sensitivity coefficients are known, BBN can be studied perturbatively. This approach is Well-known for relevant cosmological parameters, such as the neutron lifetime, number of neutrinos and baryon-to-photon ratio \cite{Pitrou20}. More recently, this has been self-consistently extended for a broad class of Grand Unified Theories \cite{BBN1,BBN2,BBN3}, where all couplings are allowed to vary.

The Lithium problem can be expressed as a statistical preference for a $70\%$ depletion of its theoretically predicted primordial (cosmological) abundance, with respect to the astrophysically measured one \cite{BBN3}, with otherwise standard physics. When the analysis is repeated for a broader class of GUT models, one finds \cite{BBN3} a mild (ca. two to three standard deviations) statistical preference for larger values of the fine-structure constant at the BBN epoch, at the parts per million level of relative variation, while the preferred Lithium depletion drops to about $65\%$. These results are qualitatively consistent across various models, although quantitatively the best-fit values of the relevant parameters do have a mild model dependence, further discussed in Ref. \citenum{BBN3}.

This means that the preference for a $\Delta\alpha/\alpha>0$ is not due to the Lithium problem, but to the aforementioned Deuterium discrepancy. A few ppm variation of $\alpha$ solves D discrepancy, given their positive correlation. Such a variation is consistent with all other currently available cosmological, astrophysical and local constraints. This helps with the Lithium problem, reducing the astrophysical depletion required, but only by a moderate amount. Thus the most likely explanation for the Lithium problem is an astrophysical one, and Ref. \citenum{BBN3} further shows that the amount of depletion needed to solve the Lithium problem can be accounted for by transport processes of chemical elements in stars---specifically, the combination of atomic diffusion, rotation and penetrative convection. On the other hand, the Helium4 abundance has relatively little statistical weight in the above analysis, given the very tight observational constraints on the primordial Deuterium abundance.

These results show that BBN is a very sensitive probe of new physics: it is quite remarkable that one can constrain the strength of the electromagnetic interaction, to parts per million level, when the universe was seconds to minutes old. Going forward, improving observed abundances of Deuterium and Helium4 by a factor of 2 to 3 will lead to stringent tests of this GUT class of models, and in particular will confirm or rule out the current preference for a $\Delta\alpha/\alpha>0$. A cosmological measurement of the Helium3 abundance (which is currently not possible) would also provide a key consistency test of the underlying physics. These will be crucial tasks for the next generation of high-resolutions ultra-stable astrophysical spectrographs.

\section{Dynamical dark energy}

Realistic models for a varying fine-structure constant usually rely on a fundamental or effective scalar field, and can be phenomenologically divided into two broad classes, dubbed Class I and Class II, for which astrophysical tests of the stability of $\alpha$ will play different roles as probes of the underlying cosmological model \cite{ROPP}.

Class I contains the models where the same degree of freedom yields dynamical dark energy and the varying $\alpha$. In this case the cosmological evolution of $\alpha$ is parametrically determined, meaning that its redshift dependence can be expressed as a function of cosmological parameters (including the matter density and those describing the dark energy equation of state) together with one or more parameter describing the coupling of the scalar field to the electromagnetic sector. In these models, constraints on $\alpha$ constrain the dark energy equation of state \cite{Pinho1,Pinho2,Pinho3}, and in particular may enable an observational discriminating test between freezing and thawing dark energy models \cite{Boas}. The simplest example of this class are quintessence models, provided one does not forget the coupling to the electromagnetic sector.

A recent example are the Euclid forecast constraints on dark energy coupled to electromagnetism, with astrophysical and laboratory data \cite{Euclid}. These show that $\alpha$ measurements improve the Euclid dark energy Figure of merit by between 8 and 26 percent, depending on the correct fiducial model (with larger improvements occurring in the null case, where the fiducial model is $\Lambda$CDM). Inter alia, these forecasts confirm the expectation \cite{Calabrese} that increasing redshift lever arm of the measurements, which the $\alpha$ data enables, is crucial.

Class II contains the models where the dynamical degree of freedom responsible for the varying $\alpha$ has a negligible effect on cosmological dynamics, and therefore has little or no contribution to the acceleration of the universe. Such models are identifiable through consistency tests comparing cosmological and astrophysical (or local) data, and in this case $\alpha$ measurements still constrain model parameters. Typically they do not directly constrain cosmological parameters, but they may still indirectly help constrain them by breaking degeneracies in the overall parameter space (which will include cosmological and particle physics parameters). The simplest example of this class are the Bekenstein type models \cite{Beken1,Beken2}. In some cases, such as the Dirac-Born-Infeld type models, astrophysical measurements of $\alpha$ may be the only possible observational probe that, given already available constraints, can distinguish these models from $\Lambda$CDM \cite{Tavares}.

Importantly, in all these models the scalar field inevitably couples to nucleons, leading to Weak Equivalence Principle violations \cite{Carroll,Damour}. Therefore measurements of $\alpha$ constrain the Eotvos parameter $\eta$. The current bound, including the available $\alpha$ measurements together with the MICROSCOPE bound \cite{Touboul}, has been reported in Ref. \citenum{Prat}, and is
\begin{equation}
    \eta<4\times10^{-15}\,.
\end{equation}
This is three times stronger than the bound from MICROSCOPE alone, and 30 time tighter than the best ground-based direct bounds, with the caveat that the constraint includes a mild model dependence (ca. factor of 2). This will be further improved by ESPRESSO, and later on by ELT-HIRES. ESPRESSO observations can probably reach a sensitivity of $2\times10^{-16}$ (about 5 times better than the final MICROSCOPE results, due to appear imminently) while the ELT-HIRES expected sensitivity is at the few times $10^{-18}$ level, similar to that of proposed STEP.

\section{So what's your point?}

The acceleration of the universe shows that canonical theories of cosmology and particle physics are incomplete, if not incorrect. Precision astrophysical spectroscopy provides a direct and competitive probe of the (still unknown) new physics that must be out there. This already provides highly competitive constraints, and will take an increasingly stronger role in the coming years.

So far one can say that nothing is varying at the few parts per million level of relative variation. This is already a very tight bound. It is orders of magnitude stronger than the few percent level of the constraints on the deviation of the dark energy equation of state from a cosmological constant, and also one order of magnitude stronger than the Cassini bound on the Eddington parameter. Tests of the stability of $\alpha$ also lead to the best available Weak Equivalence Principle constraint, a situation that is expected to remain even after MICROSCOPE's final results.

The ESPRESSO spectrograph is here, and new and more robust measurements are coming soon. Together with the final MICROSCOPE results, stringent new tests will become possible. In the longer term, the ELT will be the flagship tool in a new generation of precision consistency tests of fundamental physics, leading to competitive 'guaranteed science' implications for dark energy as well as unique synergies with other facilities, including Euclid and the SKA.

\section*{Acknowledgments}

This work was financed by FEDER---Fundo Europeu de Desenvolvimento Regional funds through the COMPETE 2020---Operational Programme for Competitiveness and Internationalisation (POCI), and by Portuguese funds through FCT - Funda\c c\~ao para a Ci\^encia e a Tecnologia in the framework of the project POCI-01-0145-FEDER-028987 and PTDC/FIS-AST/28987/2017.

\eject

\bibliographystyle{ws-procs961x669}
\bibliography{martinsconstants}

\begin{thebibliography}{10}

\bibitem{Dicke}
R.~H. Dicke, {Experimental relativity}, in {\em {Relativit\'e, Groupes et
  Topologie: Proceedings, \'Ecole d'\'et\'e de Physique Th\'eorique, Session
  XIII, Les Houches, France, Jul 1 - Aug 24, 1963}\/}, 1964.

\bibitem{Carroll}
S.~M. Carroll, {Quintessence and the rest of the world}, {\em PRL} {\bf 81},
  3067  (1998).

\bibitem{Damour}
T.~Damour and J.~F. Donoghue, {Phenomenology of the Equivalence Principle with
  Light Scalars}, {\em Class. Quant. Grav.} {\bf 27}, p. 202001  (2010).

\bibitem{ROPP}
C.~J. A.~P. Martins, {The status of varying constants: a review of the physics,
  searches and implications}, {\em Rep. Prog. Phys.} {\bf 80}, p. 126902
  (2017).

\bibitem{Meritxell}
C.~J. A.~P. Martins and M.~Vila Mi\~nana, {Consistency of local and
  astrophysical tests of the stability of fundamental constants}, {\em Phys.
  Dark Univ.} {\bf 25}, p. 100301  (2019).

\bibitem{Mortlock}
D.~J. Mortlock {\em et~al.}, {A luminous quasar at a redshift of z = 7.085},
  {\em Nature} {\bf 474}, p. 616  (2011).

\bibitem{Wilczynska}
M.~R. Wilczynska {\em et~al.}, {Four direct measurements of the fine-structure
  constant 13 billion years ago}, {\em Sci. Adv.} {\bf 6}, p. eaay9672  (2020).

\bibitem{ESPRESSO1}
F.~Pepe {\em et~al.}, {ESPRESSO at VLT - On-sky performance and first results},
  {\em Astron. Astrophys.} {\bf 645}, p. A96  (2021).

\bibitem{ESPRESSO2}
T.~M. Schmidt {\em et~al.}, {Fundamental physics with ESPRESSO: Towards an
  accurate wavelength calibration for a precision test of the fine-structure
  constant}, {\em Astron. Astrophys.} {\bf 646}, p. A144  (2021).

\bibitem{Leite}
A.~Leite, C.~Martins, P.~Molaro, D.~Corre and S.~Cristiani, {Dark energy
  constraints from ESPRESSO tests of the stability of fundamental couplings},
  {\em PRD} {\bf 94}, p. 123512  (2016).

\bibitem{Fields}
B.~D. Fields, {The primordial lithium problem}, {\em Ann. Rev. Nucl. Part.
  Sci.} {\bf 61}, 47  (2011).

\bibitem{Pitrou20}
C.~Pitrou, A.~Coc, J.-P. Uzan and E.~Vangioni, {A new tension in the
  cosmological model from primordial deuterium?}, {\em Mon. Not. Roy. Astron.
  Soc.} {\bf 502}, p. 2474  (2021).

\bibitem{BBN1}
M.~T. Clara and C.~J. A.~P. Martins, {Primordial nucleosynthesis with varying
  fundamental constants: Improved constraints and a possible solution to the
  Lithium problem}, {\em Astron. Astrophys.} {\bf 633}, p. L11  (2020).

\bibitem{BBN2}
C.~J. A.~P. Martins, {Primordial nucleosynthesis with varying fundamental
  constants: Degeneracies with cosmological parameters}, {\em Astron.
  Astrophys.} {\bf 646}, p. A47  (2021).

\bibitem{BBN3}
M.~Deal and C.~J. A.~P. Martins, {Primordial nucleosynthesis with varying
  fundamental constants: Solutions to the Lithium problem and the Deuterium
  discrepancy}, {\em Astron. Astrophys.} {\bf 653}, p. A48  (2021).

\bibitem{Pinho1}
C.~J. A.~P. Martins and A.~M.~M. Pinho, {Fine-structure constant constraints on
  dark energy}, {\em Phys. Rev. D} {\bf 91}, p. 103501  (2015).

\bibitem{Pinho2}
C.~J. A.~P. Martins, A.~M.~M. Pinho, R.~F.~C. Alves, M.~Pino, C.~I. S.~A. Rocha
  and M.~von Wietersheim, {Dark energy and Equivalence Principle constraints
  from astrophysical tests of the stability of the fine-structure constant},
  {\em JCAP} {\bf 08}, p. 047  (2015).

\bibitem{Pinho3}
C.~J. A.~P. Martins, A.~M.~M. Pinho, P.~Carreira, A.~Gusart, J.~L\'opez and
  C.~I. S.~A. Rocha, {Fine-structure constant constraints on dark energy: II.
  Extending the parameter space}, {\em Phys. Rev. D} {\bf 93}, p. 023506
  (2016).

\bibitem{Boas}
J.~M. A.~V. Boas, D.~M.~N. Magano, C.~J. A.~P. Martins, A.~Barbecho and
  C.~Serrano, {Distinguishing freezing and thawing dark energy models through
  measurements of the fine-structure constant}, {\em Astron. Astrophys.} {\bf
  635}, p. A80  (2020).

\bibitem{Euclid}
M.~Martinelli {\em et~al.}, {Euclid: constraining dark energy coupled to
  electromagnetism using astrophysical and laboratory data} (5 2021).

\bibitem{Calabrese}
E.~Calabrese, M.~Martinelli, S.~Pandolfi, V.~F. Cardone, C.~J. A.~P. Martins,
  S.~Spiro and P.~E. Vielzeuf, {Dark Energy coupling with electromagnetism as
  seen from future low-medium redshift probes}, {\em Phys. Rev. D} {\bf 89}, p.
  083509  (2014).

\bibitem{Beken1}
A.~C.~O. Leite and C.~J. A.~P. Martins, {Current and future constraints on
  Bekenstein-type models for varying couplings}, {\em Phys. Rev. D} {\bf 94},
  p. 023503  (2016).

\bibitem{Beken2}
C.~S. Alves, A.~C.~O. Leite, C.~J. A.~P. Martins, T.~A. Silva, S.~A. Berge and
  B.~S.~A. Silva, {Current and future constraints on extended Bekenstein-type
  models for a varying fine-structure constant}, {\em Phys. Rev. D} {\bf 97},
  p. 023522  (2018).

\bibitem{Tavares}
V.~C. Tavares and C.~J. A.~P. Martins, {Varying alpha generalized
  Dirac-Born-Infeld models}, {\em Phys. Rev. D} {\bf 103}, p. 023525  (2021).

\bibitem{Touboul}
P.~Touboul {\em et~al.}, {Space test of the Equivalence Principle: first
  results of the MICROSCOPE mission}, {\em Class. Quant. Grav.} {\bf 36}, p.
  225006  (2019).

\bibitem{Prat}
C.~J. A.~P. Martins and M.~Prat~Colomer, {Fine-structure constant constraints
  on late-time dark energy transitions}, {\em Phys. Lett. B} {\bf 791}, 230
  (2019).

\end{thebibliography}

\end{document}